\documentclass[preprint,showpacs,amsmath,amssymb]{revtex4}

\usepackage{graphicx}
\usepackage{dcolumn}
\usepackage{bm}

\begin{document}

\title{Evidence of $N^{*}(1535)$ resonance contribution in the $p n \to d \phi$
reaction}

\author{Xu Cao$^{1,4,6}${\footnote{Electronic address: caoxu@impcas.ac.cn}} }
\author{Ju-Jun Xie$^{2,4}${\footnote{Electronic address: xiejujun@ihep.ac.cn}} }
\author{Bing-Song Zou$^{3-6}${\footnote{Electronic address: zoubs@mail.ihep.ac.cn}} }
\author{Hu-Shan Xu$^{1,4,5}${\footnote{Electronic address: hushan@impcas.ac.cn}} }

\affiliation{$^1$Institute of Modern Physics, Chinese Academy of
Sciences, Lanzhou 730000, China\\
$^2$Department of Physics, Zhengzhou University, Zhengzhou Henan
450052, China\\
$^3$Institute of High Energy Physics, Chinese Academy of Sciences,
Beijing 100049, China\\
$^4$Theoretical Physics Center for Sciences Facilities, Chinese
Academy of Sciences,
Beijing 100049, China\\
$^5$Center of Theoretical Nuclear Physics, National
Laboratory of Heavy Ion Collisions, Lanzhou 730000, China\\
$^6$Graduate University of Chinese Academy of Sciences, Beijing
100049, China}

\begin{abstract}
The $N^{*}(1535)$ resonance contributions to the $p n \to d \phi$
reaction are evaluated in an effective Lagrangian model. The $\pi-$,
$\eta-$, and $\rho-$meson exchange are considered. It is shown that
the contributions from $\pi-$ and $\rho-$meson exchange are
dominant, while the contribution from $\eta-$meson exchange is
negligibly small. Our theoretical results reproduce the experimental
data of both total cross section and angular distribution well. This
is another evidence that the $N^*(1535)$ resonance has large
$s\bar{s}$ component leading to a large coupling to $N\phi$, which
may be the real origin of the OZI rule violation in the $\pi N$ and
$p N$ reactions.

\end{abstract}
 \maketitle{}

\section{INTRODUCTION}
The intensive interest in $\phi$-meson production in different
elementary reactions is mainly related to the investigation of the
Okubo-Zweig-Iizuka(OZI) rule violation~\cite{ozisum} which is
thought to originate from the strangeness degrees of freedom in the
nucleon and nucleon resonances. Based on the OZI rule, the ratio of
$\phi$- to $\omega$-meson production under similar kinematic
conditions are expected to be $R_{OZI}\approx tan^{2} \Delta
\theta_{V}\approx 4.2\times 10^{-3}$~\cite{ozi}, with the small
deviation $\Delta \theta_{V}=3.7^{\circ}$ from ideal mixing of octet
and singlet isoscalar vector mesons at the quark level. A
significantly apparent OZI rule violation, however, was reported in
$p\bar{p}$ annihilation at the LEAR facility at
CERN~\cite{pbardata}. Some authors attributed the origin of the OZI
rule violation to the shake-out and rearrangement of the intrinsic
$s\bar{s}$ content in the quark wave function of the
nucleon~\cite{ellis}, which was indicated by the analysis of the
$\pi$-nucleon $\sigma$-term~\cite{sigmaterm} and the lepton
deep-inelastic scattering data~\cite{dis}. This picture has also
been applied to the $\phi$-meson electro- and photoproduction off
the proton~\cite{titovstrange}, and may give a natural explanation
to the empiric evidence of a positive strangeness magnetic moment of
the proton~\cite{zou}.

Recently, OZI rule violation was found in the $pN$ collisions at the
ANKE facility at COSY~\cite{cosypphi,cosydphi}, and they obtained
$\sigma(pp \to pp\phi)/\sigma(pp \to pp \omega) = (3.3 \pm 0.6)
\times 10^{-2} \approx 8 \times R_{OZI}$~\cite{cosypphi}, and
$\sigma(pn \to d\phi)/\sigma(pn \to d\omega) = (4.0 \pm 1.9) \times
10^{-2} \approx 9 \times R_{OZI}$~\cite{cosydphi}. Several
theoretical
articles~\cite{sirpp,nakastrange,grishstrange,kaptari,kapstrange,xiephi}
were published trying to advance our understanding on this problem.
Using a relativistic meson exchange model, Nakayama et
al.~\cite{nakastrange} concluded that the mesonic current involving
the OZI rule violating $\phi\rho\pi$ vertex is dominant, while the
nucleonic current contribution had effect on the angular
distribution due to its destructive interference with the mesonic
current. They did not consider the possible role of the nucleon
resonances, because there were no experimentally observed baryonic
resonances which would decay into the $\phi N$ channel, and also the
existing data were not enough to extract the parameters relevant to
the resonances. However, the paper as well as other
ones~\cite{grishstrange,kaptari} on the $p n \to d \phi$ reaction
did not give a simultaneous good predictions to the total cross
section and angular distribution measured recently by COSY-ANKE
Collaboration~\cite{cosydphi}. In Ref.~\cite{xiephi}, it is found
that the contributions from sub-$\phi N$-threshold $N^{*}(1535)$
resonance were dominant to the near-threshold $\phi$ production in
proton-proton and $\pi^{-}p$ collisions, and all the experimental
data could be nicely reproduced by the model.

In this paper, we extend the model~\cite{xiephi} to study the $pn
\to d\phi$ reaction without introducing any further model
parameters. We assume the reaction is predominantly proceeded
through the excitation and decay of the sub-$\phi N$-threshold
$N^*(1535)$ resonance with the final nucleons merging to form the
deuteron. We calculate the total and differential cross sections of
$pn \to d \phi$ reaction in the frame of an effective Lagrangian
approach with the same value of parameters as we have well used in
Ref.~\cite{xiephi}.

Our paper is organized as follows. In Sect.~\ref{formalism}, we
present the formalism and ingredients in our computation. The
numerical results and discussion are given in
Sect.~\ref{discussion}.

\section{FORMALISM AND INGREDIENTS} \label{formalism}

\begin{figure}[htbp]
  \begin{center}
 {\includegraphics*[scale=0.8]{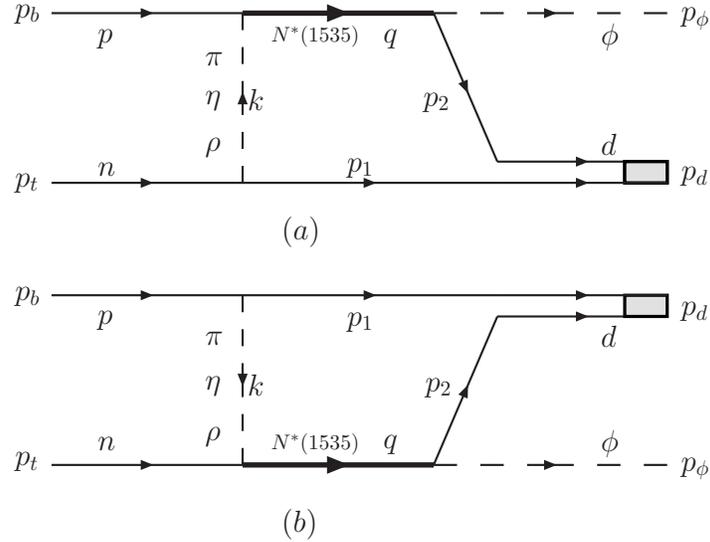}}
    \caption{Feynman diagrams for $pn \rightarrow d \phi$, (a) projectile excitation and (b) target
    excitation.} \label{fdg}
  \end{center}
\end{figure}

The Feynman diagrams for the $pn \rightarrow d\phi$ reaction are
depicted in Fig. 1, both projectile and target excitation are
included. We use the commonly used interaction Lagrangians for $\pi
NN$, $\eta NN$ and $\rho NN$ couplings,
\begin{equation}
{\cal L}_{\pi N N}  = -i g_{\pi N N} \bar{N} \gamma_5 \vec\tau \cdot
\vec\pi N, \label{pin}
\end{equation}
\begin{equation}
{\cal L}_{\eta N N}  = -i g_{\eta N N} \bar{N} \gamma_5 \eta N,
\label{etan}
\end{equation}
\begin{equation}
{\cal L}_{\rho N N} = -g_{\rho N N}
\bar{N}(\gamma_{\mu}+\frac{\kappa}{2m_N} \sigma_{\mu \nu}
\partial^{\nu})\vec\tau \cdot \vec\rho^{\mu} N. \label{rhon}
\end{equation}

At each vertex a relevant off-shell form factor is used. In our
computation, we take the same form factors as that used in the
well-known Bonn potential model~\cite{mach},
\begin{equation}
F^{NN}_M(k^2_M)=(\frac{\Lambda^2_M-m_M^2}{\Lambda^2_M- k_M^2})^n,
\end{equation}
with n=1 for $\pi$ and $\eta$-meson, and n=2 for $\rho$-meson.
$k_M$, $m_M$ and $\Lambda_M$ are the 4-momentum, mass and cut-off
parameters for the exchanged-meson ($M$), respectively. The coupling
constants  and the cutoff parameters are taken as
~\cite{xiephi,mach,tsushima,sibi}: $g_{\pi NN}^2/4\pi = 14.4$,
$g_{\eta NN}^2/4\pi = 0.4$, $g_{\rho NN}^2/4\pi = 0.9$,
$\Lambda_{\pi}$ = $\Lambda_{\eta}$ = 1.3 GeV, $\Lambda_{\rho}$ = 1.6
GeV, and $\kappa$ = 6.1.

The effective Lagrangian for $N^*(1535) N \pi$, $N^*(1535) N \eta$,
$N^*(1535) N \rho$ and $N^*(1535) N \phi$ couplings
are~\cite{xiephi},
\begin{equation}
{\cal L}_{\pi N N^*}  =  ig_{N^* N\pi}\bar{N} \vec\tau \cdot \vec\pi
N^*+h.c.,\label{pion1}
\end{equation}
\begin{equation}
 {\cal L}_{\eta N N^*} = ig_{N^* N\eta}\bar{N}\eta N^*+h.c.,
 \label{eta1}
\end{equation}
\begin{equation}
{\cal L}_{\rho N N^*} = ig_{N^* N\rho}\bar{N} \gamma_5
(\gamma_{\mu}-\frac{q_{\mu} \not \! q}{q^2}) \vec\tau \cdot
\vec\rho^{\mu} N^* +h.c.,\label{rho1}
\end{equation}
\begin{equation}
 {\cal L}_{\phi N N^*} = ig_{N^* N \phi}\bar{N} \gamma_5 (\gamma_{\mu}-\frac{q_{\mu}
\not \! q}{q^2}) \phi^{\mu} N^*+h.c..\label{phi1}
\end{equation}
Here $N$ and $N^*$ are the spin wave functions for the nucleon and
$N^*(1535)$ resonance; $\rho^{\mu}$ and $\phi^{\mu}$ are the $\rho$-
and $\phi$-meson field, respectively. For the $N^*(1535)$-$N$-Meson
vertexes, monopole form factors are used,
\begin{equation}
F^{N^* N}_M(k^2_M)=\frac{\Lambda^{*2}_M-m_M^2}{\Lambda^{*2}_M-
k_M^2}, \label{sff}
\end{equation}
with $\Lambda^*_{\pi}$ = $\Lambda^*_{\eta}$ = $\Lambda^*_{\rho}$ =
1.3 GeV.


The $N^*(1535)N\pi$, $N^*(1535)N\eta$ and $N^*(1535)N\rho$ coupling
constants are determined from the experimentally observed partial
decay widths of the $N^*(1535)$ resonance, and the coupling strength
of $N^*(1535)N\phi$ is extracted from the data of $pp\to pp\phi$ and
$\pi^- p \to n\phi$ as described in Ref.~\cite{xiephi}. For the sake
of completeness of this section, we list the values of these
parameters in Table~\ref{table}. For the $N^*(1535)N\rho$ coupling,
it is shown in Ref.~\cite{xierho} that the value is consistent with
the one estimated from the isovector radiative decay amplitude of
the $N^*(1535)$, $A^{I=1}_{1/2}=(0.068\pm 0.020)$ GeV$^{-1}$, by the
relation $A^{I=1}_{1/2}\varpropto g_{N^* N\rho}g_{\rho\gamma}$. For
the $N^*(1535)N\phi$ coupling, if the same effective Lagrangian
approach~\cite{xierho} with vector-meson-dominance is used, it can
be verified that the large value $g^2_{\phi NN^*}/4\pi = 0.13$ is
still compatible with the constraint from the small isoscalar
radiative decay amplitude of $N^*(1535)$, $A^{I=0}_{1/2}=(0.022\pm
0.020)$ GeV$^{-1}$ deduced from PDG~\cite{pdg2008}, by using the
relation $A^{I=0}_{1/2}\varpropto g_{N^*
N\omega}g_{\omega\gamma}+g_{N^* N\phi}g_{\phi\gamma}\approx (g_{N^*
N\omega}+\sqrt{2}g_{N^* N\phi})g_{\rho\gamma}/3$ and taking into
account the uncertainty of $g_{N^* N\omega}$.

\begin{table}
\caption{\label{table} Relevant $N^*(1535)$ parameters.}
\begin{center}
\begin{tabular}{|cccc|}
\hline Decay channel  & Branching ratio & Adopted branching ratio & $g^2/4 \pi$\\
\hline $N \pi$ & 0.35 -- 0.55 & 0.45 & 0.033 \\
       $N \eta$ & 0.45 -- 0.60 & 0.53 & 0.28 \\
       $N \rho \to N \pi \pi $ & 0.02 $\pm$ 0.01      &0.02  & 0.10 \\
       $N \phi$ & ---      &---      & 0.13 \\
\hline
\end{tabular}
\end{center} \label{tab1}
\end{table}

For the neutron-proton-deuteron vertex, we take the effective
interaction as~\cite{wilkin,laget},
\begin{eqnarray}
iS^{c}_{F}(p_1)(-i\Gamma_\mu\varepsilon^{\mu}_{d})iS_{F}(p_2) &=&
\frac{(2\pi)^4}{\sqrt{2}}\delta(\frac{p_d \cdot q_r}{m_d})
u(p_1,s_1) \phi_{s}(Q_R) u(p_2,s_2),
\end{eqnarray}
with $iS_{F}(p)$ being the nucleon propagator and $q_r=(p_1-p_2)/2$
the neutron-proton relative four momentum. $Q_R=\sqrt{-q_r^2}$ is
the deuteron internal momentum and $\varepsilon^{\mu}_{d}$ is the
polarization vector of the deuteron. We neglect the D-wave part of
the deuteron wave function since it gives only a minor
contribution~\cite{grishstrange}, and the S-wave deuteron wave
function $\phi_{S}(Q_R)$ can be parameterized as the Reid soft core
wave function~\cite{huthen}. We also calculate the results with
parameterized Hulth\'{e}n wave function~\cite{huthen}, which has
distinctive difference from Reid soft core wave function only below
$r=1$ fm. It gives about 20\% smaller cross section without changing
the shape of the angular distribution much and is still compatible
with available experimental data. So the different choice of the
deuteron wave function does not affect our final conclusions. But
since Reid soft core wave function is a more realistic description
of deuteron, hereafter our calculations are all based on the Reid
soft core.

Then the invariant amplitude can be obtained straightforwardly by
applying the Feynman rules to Fig. 1. Here we take explicitly the
$\pi^0$ exchange and projectile excitation diagram as an example,
\begin{eqnarray}
{\cal M}^{\pi^0,a}_{pn \to d\phi} &=& g_{\phi NN^{*}}g_{\pi
NN^{*}}g_{\pi NN}\int {\rm d}^{4}{q_r}\frac{1}{\sqrt{2}}
\delta(\frac{p_d \cdot q_r}{m_d}) \phi_{s}(Q_R)
F^{NN}_{\pi}(k_{\pi}) F^{N^*N}_{\pi}(k_{\pi})F_{N^*}(q) \times \nonumber\\
&& \bar{u}(p_2,s_2) \gamma_{5} \left(\gamma_\mu-\frac{q_\mu \not \!
q}{q^2}\right)\varepsilon^{\mu *}(p_\phi,s_\phi) G_{N^*}(q)
u(p_b,s_b)\times \nonumber\\
&& G_{\pi}(k_{\pi}) u(p_t,s_t) \gamma_{5} \bar{u}(p_1,s_1),
\label{pizero}
\end{eqnarray}
where the form factor for $N^*(1535)$ resonance, $F_{N^*}(q^2)$, is
taken as,
\begin{equation}
F_{N^*}(q^2)=\frac{\Lambda^{4}}{\Lambda^{4} +
(q^2-M^2_{N^*(1535)})^2},
\end{equation}
with $\Lambda$ = 2.0 GeV. $G_{M}(k_{M})$ and $G_{N^*(1535)}(q)$ are
the propagators of the $N^*(1535)$ resonance and exchanged meson
respectively, which can be written as~\cite{liang},
\begin{equation}
G_{\pi/\eta}(k_{\pi/\eta})=\frac{i}{k^{2}_{\pi/\eta}-m^{2}_{\pi/\eta}},
\end{equation}
\begin{equation}
G^{\mu\nu}_{\rho}(k_{\rho})=-i\frac{g^{\mu\nu}-k_{\rho}^{\mu}
k_{\rho}^{\nu}/k_{\rho}^{2}}{k^{2}_{\rho}-m^{2}_{\rho}},
\end{equation}
\begin{equation}
G_{N^*(1535)}(q)=\frac{ i(\not \! q
+M_{N^*(1535)})}{q^2-M^2_{N^*(1535)}+iM_{N^*(1535)}\Gamma_{N^*(1535)}(q^2)}.
\end{equation}
Here $\Gamma_{N^*}(q^2)$ is the energy dependent total width of the
$N^*(1535)$ resonance. According to PDG~\cite{pdg2008}, the dominant
decay channels for the $N^*(1535)$ resonance are $\pi N$ and $\eta
N$, so we take,
\begin{equation}
\Gamma_{N^*} (q^2) = \Gamma_{N^*\to N\pi}\frac{\rho_{\pi
N}(q^2)}{\rho_{\pi N}(M_{N^*}^2)}+\Gamma_{N^*\to
N\eta}\frac{\rho_{\eta N}(q^2)}{\rho_{\eta N}(M_{N^*}^2)},
\end{equation}
where $\rho_{\pi(\eta)N}(q^2)$ is the following two-body phase space
factor,
\begin{equation}
\rho_{\pi(\eta)N}(q^2)= \frac{2 p^{cm}_{\pi(\eta) N}
(q^2)}{\sqrt{q^2}} = \frac{\sqrt{(q^2-(m_N + m_{\pi(\eta)})^2)(q^2 -
(m_N-m_{\pi(\eta)})^2)}}{q^2}.
\end{equation}
It is too computer-time-consuming to directly compute
Eq.~(\ref{pizero}), and we make the same approximation as
Ref.~~\cite{laget} by ignoring the weak dependence of the dirac
spinors $\bar{u}(p_1,s_1)$ and $\bar{u}(p_2,s_2)$ to the relative
momentum $q_r$ since the deuteron wave function $\phi_{s}(Q_R)$
decreases rapidly with increasing $Q_R$. Evaluating these spinors at
the point $q_r=0$, from Eq.~(\ref{pizero}) we can straightforwardly
get the simple factorized result,
\begin{eqnarray}
{\cal M}^{\pi^0,a}_{pn \to d\phi}&=&{\cal M}^{\pi^0,a}_{pn \to
pn\phi} \times F_{\pi}(p_b,p_{\phi}),\label{pizerofa}
\end{eqnarray}
where ${\cal M}^{\pi^0,a}_{pn \to pn\phi}$ is the invariant
amplitude of process $pn \to pn\phi$ with vanishing $q_r$,
\begin{eqnarray}
{\cal M}^{\pi^0,a}_{pn \to pn\phi} &=& g_{\phi NN^{*}}g_{\pi
NN^{*}}g_{\pi NN} \bar{u}(p_2,s_2) \gamma_{5}
\left(\gamma_\mu-\frac{q_\mu \not \! q}{q^2}\right)\varepsilon^{\mu
*}(p_\phi,s_\phi) (\not \! q+M_{N^*(1535)})
u(p_b,s_b)\times \nonumber\\
&& u(p_t,s_t) \gamma_{5} \bar{u}(p_1,s_1),\label{pizerosi}
\end{eqnarray}
with $p_1=p_2=p_d/2$. On the other hand, all the four momenta in
$F_{\pi}(p_b,p_{\phi})$ are dependent on the $q_r$ and should be
integrated out,
\begin{eqnarray}
F_{\pi}(p_b,p_{\phi})=\int {\rm d}^{4}{q_r}\frac{1}{\sqrt{2}}
\delta(\frac{p_d \cdot q_r}{m_d}) \phi_{s}(Q_R)
\frac{F^{NN}_{\pi}(k_{\pi}) F^{N^*N}_{\pi}(k_{\pi}) F_{N^*}(q)
G_{\pi}(k_{\pi})}{q^2-M^2_{N^*(1535)}+iM_{N^*(1535)}\Gamma_{N^*(1535)}(q^2)},
\end{eqnarray}
This prescription could largely reduce the laborious computation,
and a comparison of the full calculation Eq.~(\ref{pizero}) and the
approximation Eq.~(\ref{pizerosi}) will be given later. Diagrams for
the target excitation and other exchanged mesons are in the similar
fashion. Isospin factors should be considered to take into account
the contribution of charged mesons. Then the differential and total
cross sections are calculated by,
\begin{eqnarray}
\frac{d\sigma}{d\Omega} = \frac{m_p m_d m_n}{8\pi ^2 s}
\frac{|\vec{p_\phi}|}{|\vec{p_t}|} \sum_s |{\cal M}_{pn\rightarrow
d\phi}|^2.
\end{eqnarray}
with ${\cal M}_{pn\rightarrow
d\phi}=\sum\limits_{i=\pi,\eta,\rho}({\cal M}^{i,a}_{pn\rightarrow
d\phi}+{\cal M}^{i,b}_{pn\rightarrow d\phi})$. The interference
terms are ignored in our concrete calculations because the relative
phases among different meson exchanges are unknown.

\section{Numerical RESULTS AND DISCUSSION} \label{discussion}

\begin{figure}[htbp]
  \begin{center}
{\includegraphics*[scale=1.0]{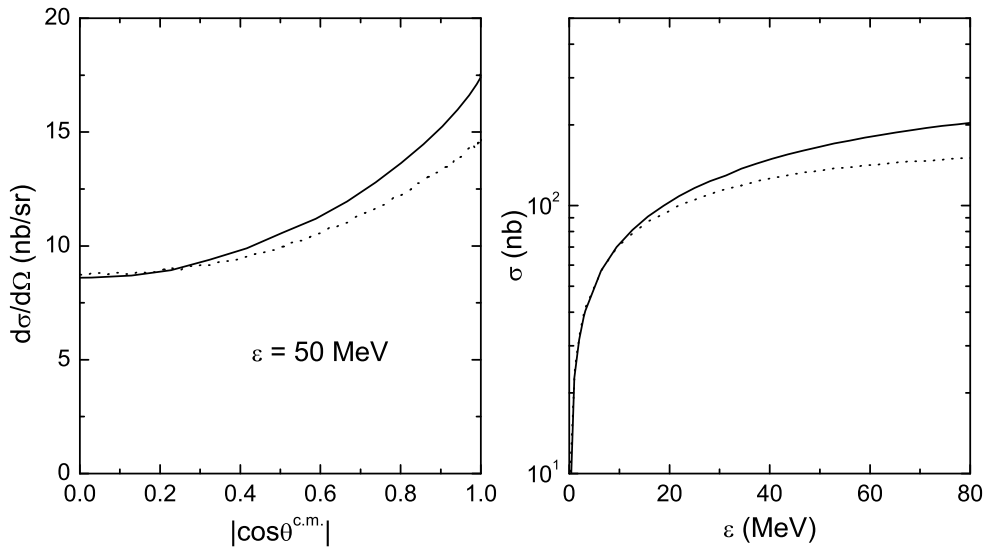}}
    \caption{$\pi$-meson exchange contribution to
the cross section (Right) and $\phi$-meson polar angular
distribution in excess energy 50MeV (Left). Solid lines represent
the full calculation of Eq.~(\ref{pizero}), and dotted lines are the
results of the calculation with approximation of
Eqs.~(\ref{pizerofa})~(\ref{pizerosi}).} \label{compare}
  \end{center}
\end{figure}

Fig.~\ref{compare} shows the $\pi$-meson exchange contribution to
the cross section and $\phi$-meson polar angular distribution in
excess energy 50MeV. The difference of the full calculation
Eq.~(\ref{pizero}) and the approximation Eq.~(\ref{pizerosi}) is
tolerable, and the former gives a slightly deeper rise in the
angular distribution. Obviously this will not affect our final
conclusions, so we will confidently use the approximation in our
following calculations.

\begin{figure}[htbp]
  \begin{center}
{\includegraphics*[scale=1.0]{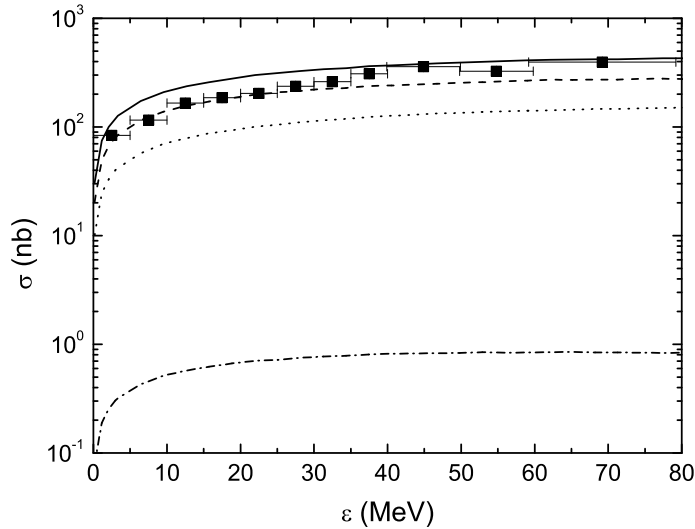}}
    \caption{Total cross section for $pn \to d \phi$. The dotted,
    dashed, dash-dotted and solid curve correspond to contribution from
    $\pi-$, $\rho-$, $\eta-$meson exchange and their simple sum, respectively.
    The data are from Ref.\protect\cite{cosydphi}.} \label{tcs}
  \end{center}
\end{figure}

With the formalism and ingredients given above, the total cross
section versus excess energy $\varepsilon$ is calculated by the
parameters fixed in the previous study~\cite{xiephi}. Our numerical
results are depicted in Fig.~\ref{tcs} together with the
experimental data. The dotted, dashed, dash-dotted and solid curve
correspond to contribution from $\pi-$, $\rho-$, $\eta-$meson
exchange and their simple incoherent sum, respectively. In the
calculation~\cite{xiephi} of $pp \to pp \phi$ reaction, contribution
from the $\pi$-meson exchange is larger than that from the
$\rho$-meson exchange by a factor of 2. Contrarily, in
Fig.~\ref{tcs}, we can see that $\rho$-meson exchange is larger than
$\pi$-meson exchange by a factor 2 in $pn \to d\phi$ reaction in the
present calculation. The main reason is that the use of deuteron
wave function for the $pn$ final state interaction gives an
enhancement factor to the $\rho$ exchange diagram about a factor of
4 larger than to the $\pi$ exchange compared with results without
including any $pn$ FSI. In the calculation~\cite{xiephi} of $pp \to
pp \phi$ reaction, a simple global Jost factor is used for the $pp$
FSI as many other previous calculations, and gives an equal
enhancement factor to all meson exchanges. As pointed out by
Ref.~\cite{wilkin}, this kind of treatment of FSI seems too simple.
For the $pp\to pp\eta$, the use of Paris wave function for the NN
FSI results in enhancement factor about a factor of 1.75 larger for
the $\rho$ exchange than for the $\pi$ exchange. In our present
calculation of $pn \to d\phi$ reaction with $pn$ as a bound state,
the enhancement factor is then understandably more larger for $\rho$
exchange than for the $\pi$ exchange. The contribution from
$\eta$-meson exchange is about three orders of magnitude smaller
than that of $\rho$-meson exchange. This relative magnitude is
smaller compared to the case of $pp \to pp \phi$ reaction. The
relative suppression of $\eta$-meson exchange is due to its
iso-scalar property while the iso-vector mesons play more important
role in the $pn$ interaction due to participation of their charged
members. The simple incoherent sum of these contributions can give a
nice description of the experimental data.

\begin{figure}[htbp]
  \begin{center}
 {\includegraphics*[scale=1.5]{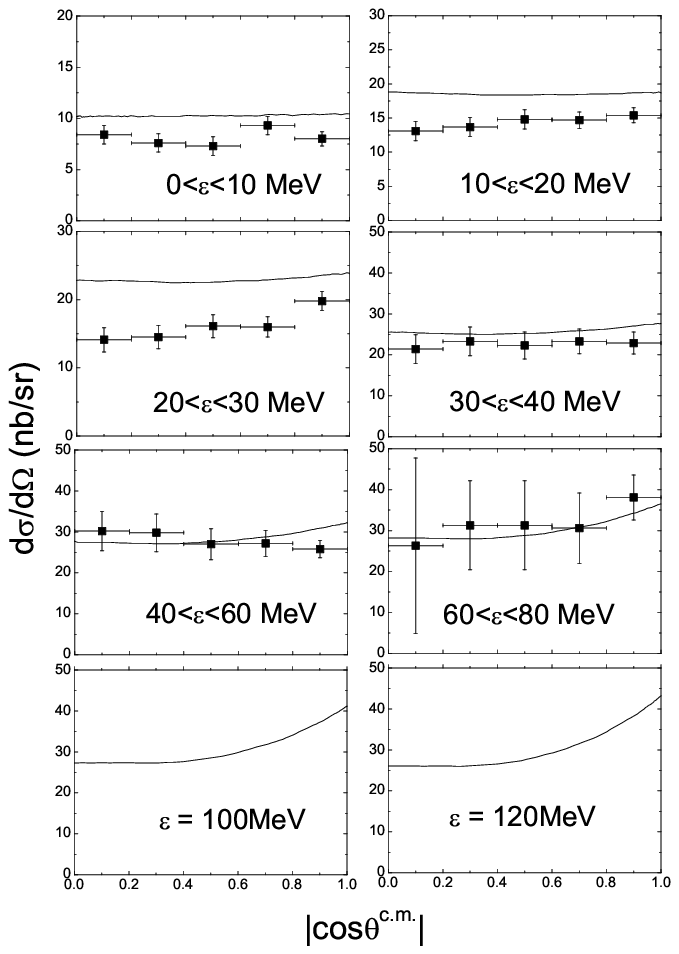}}
    \caption{Angular distributions of $\phi$ meson polar angular in the overall c.m. system.
    The data are from Ref.~\protect\cite{cosydphi}.} \label{dcs}
  \end{center}
\end{figure}

As shown in Fig.~\ref{dcs}, our calculated $\phi$-meson polar
angular distributions are compatible with the experimental data and
show some structure in high excess energy. It is seen that our
angular distributions of $pn\rightarrow d\phi$ follow the behaviour
of the corresponding distributions in $pp \rightarrow pp \phi$
reaction, modified slightly by the neutron-proton-deuteron vertex.
The upward bending at forward and backward angles becomes more
pronounced with the increasing excess energy, and it would be
possible for the experiment performed in higher energies to verify
these structures.

There are some interesting findings if we compare our results with
those of others. In the model of Nakayama et al.~\cite{nakastrange},
only mesonic and nucleonic current were considered, and they claimed
that it was necessary to introduce an OZI rule violation at the
$\phi\rho\pi$ vertex in the mesonic current, which provided the
enhancement of the $\phi$-meson production. Four parameter sets
extracted from the analysis of $pp \to pp \phi/\omega$ were used to
study the $pn \to d\phi$ reaction, but none of them could give a
simultaneous explanation to the experimental data. The model
parameter sets 1 and 2 underestimated the total cross section
slightly though they can give a fairly flat angular distributions up
to excess energy 100 MeV. The sets 3 and 4 reproduced much better
the total cross section but the predicted angular distributions
showed obvious downward bending at forward and backward angles,
which was somewhat inconsistent with the experimental data. Those
characteristics might mean that it could not reasonably account for
the reaction dynamics of the $pn \to d\phi$ reaction by only
including mesonic and nucleonic currents. Kaptari et
al.~\cite{kaptari} used a modified model including the
bremsstrahlung and conversion diagrams, corresponding to the
nucleonic and mesonic currents respectively, and found conversion
diagrams were predominant without introducing obvious OZI violation
in $\phi\rho\pi$ vertex. They predicted a rather small total cross
section though their angular distribution results in the
near-threshold region seemed to be consistent with the experimental
data. Another theoretical work was finished by Grishina et
al.~\cite{grishstrange}, and their two-step model slightly
underestimated the total cross section, though this might be
attributed to the adopted large normalization factor arising from
the initial state interaction. This normalization factor seemed to
be somewhat arbitrary and it was a pity that they did not give their
angular distributions. As clearly illustrated in Fig.~\ref{tcs} and
Fig.~\ref{dcs} as well as Ref.~\cite{xiephi}, if the $N^{*}(1535)$
resonance is dominant in the $\phi$ production in nucleon-nucleon
collisions, a consistent description of $pp \to pp\phi$ and $pn \to
d\phi$ reactions can be acquired. Certainly, it has to be admitted
that it cannot definitely exclude the contribution from the mesonic
and nucleonic current because alternative combination of those
currents and $N^{*}(1535)$ resonance would yield a good fit to the
present data. Especially, it is noted that $N^{*}(1535)$ resonance
gives upward bending but those currents give downward bending at
forward and backward angles, and their merging is expected to give
much flatter angular distributions as present data have shown. The
higher energy data should be helpful to decide the portion of these
contributions since the bending behavior is more prominent for the
excess energy above 100MeV.

According to above analysis, it is safe to conclude that the
contribution from $N^{*}(1535)$ resonance plays important role for
the $\phi$-meson production in $pN$ collisions and may be the real
origin of the large OZI rule violation. The significant
$N^{*}(1535)N\phi$ coupling alone would be enough to explain the
enhancement in the $\phi$-meson production in $pN$ collisions, and
this may indicate large $s\bar{s}$ component in quark wave function
of $N^{*}(1535)$ resonance and hence the large coupling of
$N^{*}(1535)$ to strangeness decay channels~\cite{xiephi,caoeta}.

The large $N^{*}(1535)N\phi$ coupling should also play important
role in other relevant processes. In the study of the $\phi$-meson
production in the $\bar pp$ annihilations, the strange hadron loops,
such as $K\bar{K}$, $K^*\bar{K}$, $\Lambda\bar{\Lambda}$ loops, are
found to play important role~\cite{rescatter}. It would be
interesting to investigate the contribution through $N^*(1535)$ and
$\bar N^*(1535)$ excitations. For the $\pi N\to\phi N$ reaction,
although the total cross sections can be reproduced by the t-channel
$\rho$ exchange and/or subthreshold nucleon pole
contributions~\cite{sirpp,titov}, these contributions are very
sensitive to the choice of off-shell form factors for the t-channel
$\rho$ exchange and the $g_{NN\phi}$ couplings and can be reduced by
orders of magnitude within uncertainties of these ingredients.
Alternative mechanisms~\cite{xiephi,doring} with large $N^*(1535)$
contribution can reproduce data perfectly. For the $\gamma p \to\phi
p$ reaction, a much larger OZI rule violation for $\phi$-meson
production was suggested~\cite{sirphoto,phiphoto,phiphoto2} with no
indications for s-channel resonances above
threshold~\cite{phiphoto2}. The t-channel diffractive Pomeron
exchange with photon transition to $\phi$ is found to play dominant
role~\cite{zhaotitov,titov}, but further mechanisms are needed to
account for the bump structure in the forward angle differential
cross section at low energy region~\cite{phiphoto}. It would be
interesting to check the role of $N^*(1535)$ and/or other s-channel
$N^*$ through polarization observables. The role of $N^*(1535)$ can
also be further explored in the $pd\to ^{3}$He$\phi$
reaction~\cite{hedata}, though this channel is convoluted with the
large momentum transfer between the deuteron and $^{3}$He. A
two-step model~\cite{wilkinphi} underpredicted the total cross
section by at least a factor of four, and the reaction dynamics
involving $N^{*}(1535)$ resonance may be necessary to resolve the
$\phi$ production mechanism in this reaction.

In summary, we have phenomenologically investigated the role of the
$N^{*}(1535)$ resonance in $pn\rightarrow d\phi$ reaction near
threshold, and all model parameters are taken from a previous study
of the $pp \rightarrow pp \phi$ reaction~\cite{xiephi}. We have
shown that the including of the dominant $N^{*}(1535)$ resonance
contribution is necessary to reproduce the recently measured total
and differential cross sections, though mesonic and nucleonic
currents might also have some minor contributions. We argue that the
large coupling of the intermediate $N^{*}(1535)$ resonance to
$\phi$-meson maybe an very important origin of the OZI rule
violation in the $\phi$-meson production. This can be further
investigated in various other relevant reactions.

\begin{acknowledgments}

We are grateful to C. Wilkin,  D. Y. Chen and Z. Ouyang for useful
discussion. This work was supported by Chinese Academy of Sciences
Knowledge Innovation Project (Nos. KJCX3-SYW-N2, KJCX2-SW-N16) and
the National Natural Science Foundation of China (Nos. 10847159,
10875133, 10821063, 10635080, 0701180GJ0).

\end{acknowledgments}

\end{document}